\documentclass[superscriptaddress,onecolumn,preprint ]{revtex4-1}
\usepackage{graphicx}
\usepackage{amssymb}
\usepackage{epstopdf}
\usepackage{amsmath,dsfont, commath}
\usepackage{bm}
\usepackage{braket}
\usepackage{changes}
\newcommand{\be}{\begin{equation}}
	\newcommand{\ee}{\end{equation}}
\newcommand{\bea}{\begin{eqnarray}}
	\newcommand{\eea}{\end{eqnarray}}

\newcommand{\ba}{\begin{array}}
	\newcommand{\ea}{\end{array}}

\newcommand{\bl}{\begin{flalign}}
	\newcommand{\enl}{\end{flalign}}

\newcommand{\mc}[1]{\mathcal{#1}}

\newcommand{\eq}[1]{Eq. \eqref{#1}}

\newcommand{\Eq}[1]{Equation \eqref{#1}}
\newcommand{\fig}[1]{Fig. (\ref{#1})}

\usepackage[capitalise]{cleveref}

\newcommand{\tord}{\mathcal{T}}

\newcommand{\tr}{\text{Tr}}

\makeatletter
\newsavebox{\@brx}
\newcommand{\llangle}[1][]{\savebox{\@brx}{\(\m@th{#1\langle}\)}%
	\mathopen{\copy\@brx\kern-0.5\wd\@brx\usebox{\@brx}}}
\newcommand{\rrangle}[1][]{\savebox{\@brx}{\(\m@th{#1\rangle}\)}%
	\mathclose{\copy\@brx\kern-0.5\wd\@brx\usebox{\@brx}}}
\makeatother

\begin{document}
	
	\title{Diagrammatic representation and nonperturbative approximation of exact time-convolutionless master equation }
	\author{Bing Gu}
	\email{gubing@westlake.edu.cn}
	\affiliation{
		Department of Chemistry \& Department of Physics, Westlake University, Hangzhou, Zhejiang 310030, China}
	\affiliation{Institute of Natural Sciences, Westlake Institute for Advanced Study, Hangzhou, Zhejiang 310024, China}

\begin{abstract}
	The time-convolutionless master equation provides a general framework to model non-Markovian dynamics of an open quantum system with a time-local generator. 
A diagrammatic representation is developed and proven for the perturbative expansion of the exact time-local generator for an open quantum system interacting with arbitrary environments. A truncation of the perturbation expansion leads to the perturbative time-convolutionless quantum master equations. We further introduce a nonperturbative approach that approximates the time-convolutionless generator as a nested time-ordered exponential function.
\end{abstract}
	\maketitle
	
		\section{Introduction}
	
	Quantum systems  inevitably interacts with surrounding environments that leads to quantum decoherence and dissipation \cite{breuer2002, Joos2013, schlosshauer2007, gu2017, gu2018}. Understanding  environmental influences to the system dynamics is vital to  a wide variety of problems in physics and chemistry such as quantum information processing \cite{Nielsen2011}, linear and nonlinear spectroscopy \cite{mukamel1995, gu2020c}, quantum transport \cite{jin2008}, quantum-classical transition \cite{schlosshauer2007}, quantum measurements, electron and energy transfer in condensed phase \cite{nitzan2006, scholes2017} and quantum control \cite{shapiro2003}. 
	
	As the complete description of the total system plus environment is intractable due to the large number of degrees of freedom, the common strategy is to  focus on  the primary system of interest. By tracing out the bath degrees of freedom, 
	a quantum master equation can be obtained for the reduced density matrix of the system.  
Markovian quantum master equations such as  the Lindblad and Redfield equations are widely used. They are derived under the Born approximation that only includes the leading second-order system-bath coupling and the Markov approximation that neglects the memory effects. The Redfield equation can be expressed in the Lindblad form if the secular approximation is further invoked.
 The Markovian quantum master equations all take a time-local form, i.e., \be \dot{\rho}(t) = \mc{L} \rho(t)\ee 
	where $\mc{L}$ is a Liouvillian that accounts for the environment-induced decoherence and dissipation. 
For environments with structured spectral density and strong system-bath coupling, the memory effects can be important \cite{breuer2016}. 
 
	 To go beyond the Markovian approximation, there are two class of approaches. The first is based on the time-nonlocal Nakajima-Zwanzig master equation \cite{smirne2010, xu2018}, which involves a time-convolution of the memory kernel and  history of the system. Alternatively, one can use the  time-convolutionless master equation \cite{breuer2001, kidon2015}, in which the generator of time-translation is time-local 
	 but without invoking the Markovian approximation. The memory effects is contained in the time-dependence of the time-local generator. By  contrast, the generator in Markovian master equations is  time-independent.  
	 		For Gaussian environments such as non-interacting bosons and fermions, exploiting the Gaussian statistics, there are numerically exact methods including the hierarchical equation of motion and quasi-adiabatic propagator path integral methods \cite{tanimura2020, makri1998, yan2014}. 
The time-convolutionless master equation is usually derived using the projection operator technique, but this involves super-operators defined in the full system-bath Hilbert space \cite{timm2011}.
  
			 We previously devised a set of diagrammatic rules  to represent the perturbative expansion of the time-local generator \cite{gu2020e}   for an arbitrary  open quantum system.
	Here we first  provide  a proof for the   diagrammatic representation of the exact time-local quantum master equation. This is proven by deriving a recursive equation for the perturbation series of the time-local generators from the diagrammatic rules. A $n$th-order  perturbative time-convolutionless master equation is thus obtained by retaining only terms up to the $n$th order of the generator.
	 Non-perturbative theories are required if the system-bath coupling becomes strong.  To go beyond the perturbative treatment of system-bath coupling, we introduce a non-perturbative approximation to the generator, which can be useful to treat strongly coupled system-bath models.   
	
	This paper is structured as follows. In \cref{sec:rec}, we first recap the diagrammatic representation of the time-convolutionless generator followed by deriving a recursive equation for the expansion terms. We introduce the nonperturbative approximation for the generator in \cref{sec:approx}. \cref{sec:summary} summaries.

	\section{Diagrammatics of the generator} \label{sec:rec}
	Consider a general system-bath Hamiltonian
	\be H = H_\text{S} + H_\text{B} + H_\text{SB}
	\ee  
	where $H_\text{S}$ describes the primary system of interest, $H_\text{B}$ the bath, and $H_\text{SB} $ the system-bath interaction. The most general form for the system-bath interaction reads $H_\text{SB} = \sum_i S_i \otimes B_i$, where $S_i, B_i$ are respectively, the system and bath operators.  The tensor product notation is suppressed in the following.  Hereafter, we  consider a single interaction term $H_\text{SB} = S B$ and time-independent system and bath Hamiltonian.   Generalizations to multiple interaction terms 
	and the time-dependent system and bath Hamiltonians is  straightforward.  
	Note that the partition of a total system into the primary system and bath is arbitrary.  
	
	For an initially uncorrelated state $\rho_0 = \rho_\text{S}(t_0) \rho_\text{B}(t_0)$, the exact time-local quantum master equation for the system reduced density matrix $\rho_\text{S}(t)$ in the interaction picture of $H_0 = H_S +H_B$ can be formally written as
	\be
	\dot{\rho}_\text{S}(t) = 
	\mc{G}(t) \rho_\text{S}(t)
	\label{eq:qme}
	\ee
	where  $\mc{G}(t)$ is the generator of time translation encoding all the influences of the environment including relaxation and quantum decoherence.  \Eq{eq:qme} can be  understood as the definition of the generator. 
	
	To find $\mc{G}$, we start from the Liouville equation for the total density matrix of the full system in the interaction picture of $H_0$, 
	\be
	\dot{\rho}_{t} = -i  \sum_{\sigma = \pm} \mc{S}^\sigma(t) \mc{B}^{\bar{\sigma}}(t)  \rho_{t}  
	\label{eq:110}
	\ee 
	where $\bar{\sigma} = - \sigma$. 
	  In \eq{eq:110}, 
	  we have made use of the  identity 
	\be
	[S(t)B(t), \rho] = \sum_{\sigma = \pm} \mc{S}^\sigma(t) \mc{B}^{\bar{\sigma}}(t) \rho
	\ee
	where $\mc{S}^\pm \rho = \frac{1}{\sqrt{2}} [S, \rho]_\pm = \frac{1}{\sqrt{2}} \del{S\rho \pm \rho S}$ are superoperators of commutator and anticommutator. 
	Integrating \eq{eq:110}	from $t_0$ to $t$ yields 
	\be 
		{\rho}_t  = \rho_0  -i \sum_{\sigma = \pm} \int_{t_0}^t \dif t' \mc{S}^\sigma(t') \mc{B}^{\bar{\sigma}}(t')  \rho_{t'}.  
		\label{eq:111}
	\ee 
	Inserting \eq{eq:111} into the right hand side of itself recursively,  and traceing out the bath degrees of freedom yields 
	$\rho_\text{S}(t) = \Lambda_{t, t_0} \rho_\text{S}(0)$, with the dynamical map given by a perturbation series 
	\begin{equation}
		\begin{split} 
			\Lambda_{t, t_0} &= \braket{\tord e^{-i \int_0^t \dif t' \mc{S}^\sigma(t') \mc{B}^{\bar{\sigma}}(t') } }_\text{B} \\
			&=  1 + \sum_{n=1}^\infty \sum_{\bm \sigma} \del{-i}^n  \int_{t_n > \cdots > t_1} \dif t_n \cdots \dif t_1 \mc{S}^{\sigma_n}(t_n)  \cdots \mc{S}^{\sigma_1}(t_1)   \Braket{\mc{B}^{\bar{\sigma}_n}(t_n)  \cdots \mc{B}^{\bar{\sigma}_1}(t_1) }_\text{B} \\
						& \equiv \sum_{n=1}^\infty \mc{M}^{(n)}(t) 
		\end{split}
		\label{eq:113}
	\end{equation}
	where $\tord$ denotes the Dyson time-ordering operator, $\mc{M}^{(n)}(t)$ are the perturbative expansion of the dynamical map, and $\braket{\cdots }_\text{B} = \tr\cbr{\cdots \rho_\text{B}(t_0) }$.
	
	The time-local generator can then be obtained by  
	\be
	\mc{G} = \dot{\Lambda}_{t, t_0}\Lambda_{t, t_0}^{-1} 
	\label{eq:def}
	\ee 
	Inserting \eq{eq:113} into \eq{eq:def} and	matching terms order by order of the system-bath coupling can yield the generator $\mc{G}^{\del{n}}$. However, this process quickly becomes laborious with order $n$ and the diagrammatic rules drastically simplifies this step.

	A diagrammatic analysis  has been proposed for the perturbative series  of the generator \cite{gu2020e} for general system-bath models, 
	 i.e., 
	 \be 
	 \mc{G}(t) = \sum_{n=1}^\infty \mc{G}^{(n)}(t). 
	 \label{eq:112}
	  \ee 
	  where we have taken into account that $\mc{G}^{(0)} = 0$. 
	  The diagrammatic rules are  summarized blow with slight modifications.  
	%
	The $n$-th order generator $\mc{G}^{\del{n}} $ is represented by a connected diagram with $n$ vertices together with all possible disconnected diagrams that can be obtained by  cutting the links (or ``bonds'') of the connected diagram. Each cut introduces a minus sign. Equivalent diagrams only count once.

	For example, the second-order term can be read off from \fig{fig:2nd}
\be
\mc{G}^{(2)} (t) = \del{-i}^2 \sum_{\sigma_1 = \pm} \int_0^t \dif t_1 \mc{S}^-(t) \mc{S}^{\sigma_1}\del{t_1} \braket{\mc{B}^+(t) \mc{B}^{\bar{\sigma}_1}(t_1)   } + i \mc{S}^-(t) \braket{\mc{B}(t)}_\text{B}
\ee 
The second term can be absorbed into the system Hamiltonian.

	\subsection{Recursive equation from the diagrammatics}
	We now prove that the diagrammatic rules  provide a faithful representation for the generators.
	The idea is to derive the recursive equation satisfied by $\mc{G}^{(n)}$ from the diagrammatics, 
	 and compare it with the existing one \cite{gasbarri2018}. 
	
\begin{figure}
	\centering
	\includegraphics[width=0.2\linewidth]{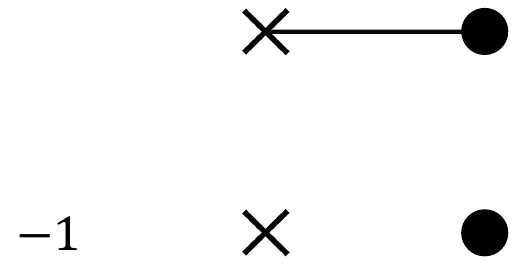}
	\caption{Diagrams for the second-order generator $\mc{G}^{(2)}$. The $-1$ in the second line signifies the minus sign associated with a single cut. }
	\label{fig:2nd}
\end{figure}
\begin{figure}
	\centering
	\includegraphics[width=0.7\linewidth]{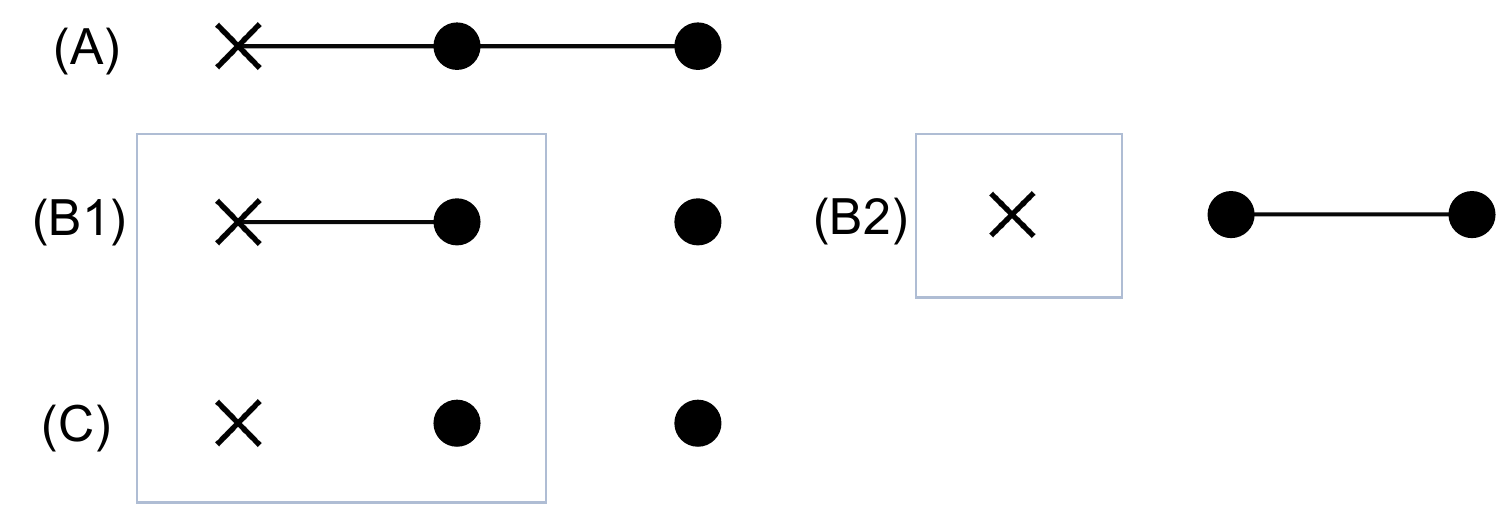}
	\caption{Diagrams illustrating the recursive equation \eq{eq:recursive} for $n = 3$.}
	\label{fig:3rd}
\end{figure}

	Here we use diagrams to derive the recursive equation.  All  diagrams contributing to the third-order time-local are shown in \fig{fig:3rd}, 
	\be
	\mc{G}^{(3)} = A - \del{B_1 + B_2 } + C =  A - \del{B_1 - C} - B_2
	\label{eq:101}
	\ee 
	There is a minus sign in front of the B diagrams because it is obtained by  cutting once  the A diagram. 
	We have simply rearranged the diagrams in the last step of \eq{eq:101}. The first term represented by diagram A is simply the time-derivative of the third-order moment $\dot{\mc{M}}^{(3)}$. 
	The second term $B_1 -C$  is in fact the second-order generator multiplied by the first-order moment, i.e., $\mc{G}^{(2)}  \mc{M}^{(1)}$. The third term is the first-order generator multiplied by the first-order moment 
	$ \mc{G}^{(1)} \mc{M}^{(2)}$, as can be seen in \fig{fig:2nd}. Thus, 
	\be
	\mc{G}^{(3)} = \dot{\mc{M}}^{(3)} - \mc{G}^{(2)}\mc{M}^{(1)} - \mc{G}^{(1)}  \mc{M}^{(2)}
	\ee


%
%

\begin{figure}
	\centering
	\includegraphics[width=\linewidth]{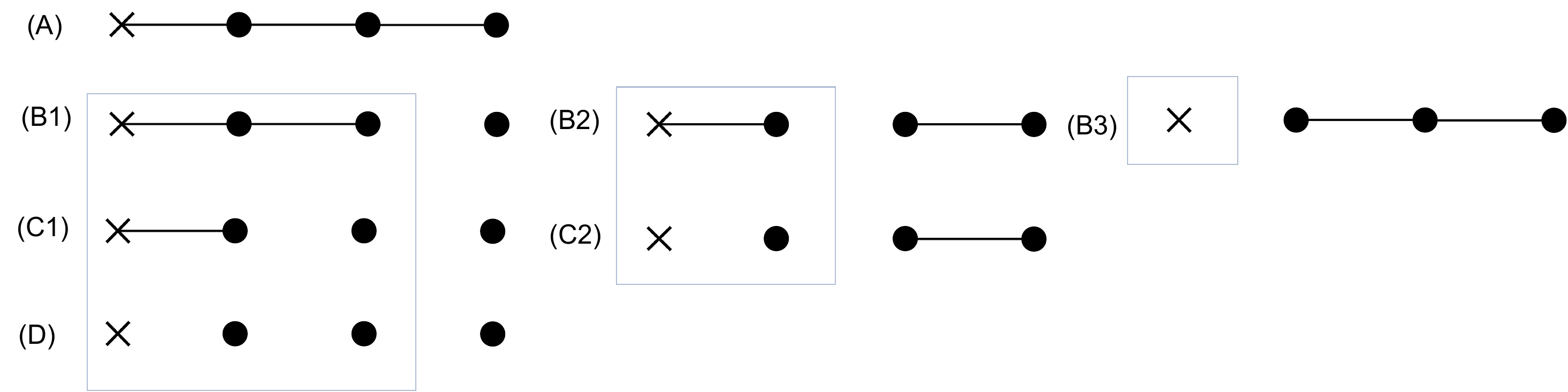}
	\caption{Diagrams illustrating the recursive equation \cref{eq:105}.}
	\label{fig:4th}
\end{figure}

The fourth-order term can be analyzed similarly. The diagrams are shown in \fig{fig:4th}. 
In terms of the diagrams 
\be
\mc{G}^{(4)} = A -  B_1 - B_2 - B_3 + C_1 + C_2 - D = A - \del{B_1 - C_1 + D} - (B_2 - C_2) - B_3 
\label{eq:102}
\ee 
In the right-hand side of \eq{eq:102}, the first term is the time-derivative of the four-order moment, $\dot{\mc{M}}^{(4)}$. The second term is simply $\mc{G}^{(3)}$ multiplied by  the first-order moment. The second is the second-order generator (see \fig{fig:2nd}) multiplied by the second-order moment. The third term is the second order generator multiplied by the second-order moment; and the last term is the first-order generator multiplied by the third-order moment.
Thus, 
\be
\mc{G}^{(4)} = \dot{\mc{M}}^{(4)} -  {\mc{G}}^{(3)} \mc{M}^{(1)} -  {\mc{G}}^{(2)} \mc{M}^{(2)} - {\mc{G}}^{(1)} \mc{M}^{(3)} 
\label{eq:105}
\ee 
The higher-order terms can be analyzed in the exactly same way. which then leads to 
a recursive relation  
\be
\mc{G}^{\del{n}} = \dot{\mc{M}}^{(n)} - \sum_{m=1}^{n-1} \mc{G}^{\del{m}} \mc{M}^{(n-m)}
\label{eq:recursive}
\ee
The equivalence of \eq{eq:recursive}  with the recursive equation in Ref. \cite{gasbarri2018} thus completes the proof of the diagrammatic representation. 

\subsection{Gaussian environments}
The discussion so far are general for an arbitrary open quantum system. Note that the bipartition of the full Hamiltonian is arbitrary,   and the system and bath can exchange roles, with the time-local generator remaining the same form\cite{gu2020e}. Specializing to the case where  the bath operator satisfies Gaussian statistics, i.e., the bath correlation functions are completely determined by the two-point correlation functions through Wick's theorem 
\be
D(1,2,\cdots, {2m}) = \sum_{\pi \in \text{pairs}} D(\pi_1, \pi_{2})D(\pi_3,\pi_4)\cdots D(\pi_{2m-1},\pi_{2m})   
\ee 
The Gaussian environment  are widely used in chemistry and physics, e.g., the spin-boson model. It can be exact for photon and phonon environments.
For Gaussian environments, the dynamical map in \eq{eq:113} can be further simplified to yield the dynamical map 
\be
\begin{split} 
	\Lambda_{t, t_0} &= {\tord \exp\del{- \iint_0^t \dif t_2 \dif t_1 \mc{S}^{\sigma_2}(t_2) \mc{S}^{\sigma_1}(t_1)    D_{\bar{\sigma}_2 \bar{\sigma}_1}(t_2, t_1)} } 
\end{split}
\label{eq:114}
\ee 
where $D_{\sigma_2 \sigma_1}(t_2, t_1) = \braket{ \tord \mc{B}^{\sigma_2}(t_2) \mc{B}^{\sigma_1}(t_1) }_\text{B}$.    Taking the time derivative of the Gaussian dynamical map repeatedly gives rise to a hierarchy of equations of motion, which, terminated appropriately, can lead to numerically exact solutions of the reduced system dynamics \cite{tanimura2020}.

%
%
%
%

\section{non-perturbative approximations} \label{sec:approx}

By only retaining the low-order expansions yields a quantum master equation in the time-local form. Yet, we expect this to be valid when the system-bath coupling is weak. For strong system-bath coupling, non-perturbative approaches are needed.  Although numerically exact methods have been developed for Gaussian environments which fully exploits the fact that a two-point correlation function completely characterizes the statistics of the environment. For non-Gaussian environments, higher-order correlation function cannot be expressed by lower-order ones, thus approximations are inevitable.

Here we show how non-perturbative approximations can be obtained by iterating the procedure that takes us from the dynamical map $\Lambda$ to the time-local generator $\mc{G}$. In this approximation, the time-local structure is retained. 
The idea of time-convolutionless master equation can be understood as that we, instead of directly approximating the dynamical map, approximate its time-local generator. This process can be iterated, that is, instead of approximating the generator, we can approximate the generator  for the generator itself. 

%
To proceed, left-multiplying $[\mc{G}^{(1)}]^{-1}$ to \eq{eq:114} yields 
\be
\tilde{\mc G} \equiv [\mc{G}^{(1)}]^{-1} \mc G 
 =1 +  \sum_{n=2}^\infty \tilde{\mc G}^{(n)}(t)
\ee  
where $\tilde{\mc  G}^{(n)} = \sbr{\mc G^{(1)}}^{-1} \mc G^{(n)}$.
In cases where the low-order terms vanish, we simply left-multiply the lowest-order non-vanishing element. If $\mc{G}^{(1)} = 0$ , we left-multiply $[\mc{G}^{(2)}]^{-1}$. 
The time-local generator for  $\tilde{\mc G}$, defined by $ \mc{G}_2 \equiv \dot{\tilde{\mc G}} \tilde{\mc G}^{-1}  $, is then 
\be 
\mc{G}_2 = \del{\sum_{n=2}^\infty \dot{\tilde{\mc G}}^{(n)} } \del{ 1 + {\sum_{m=2}^\infty   
		\tilde{\mc G}^{(m)}	}  }^{-1}  = \sum_{n=1}^\infty \mc G_2^{(n)}.
\ee 
Thus, 
\be
\mc{G}(t) = \mc{G}^{(1)}(t) \tord \exp\del{ \int_{t_0}^t \dif s \mc{G}_2(s) } 
\label{eq:104}
\ee 
By iterating this procedure, we can obtain a hierarchy of time-local generator for $\mc{G}_n(t)$, 
\be
\dot{\tilde{\mc{G}}}_n(t)  = \mc{G}_{n+1}(t) \tilde{\mc{G}}_n(t) ,
\ee
where $\mc{G}_{n+1}$ is the time-local generator for $\mc{\tilde{G}}_n(t)$ and $\mc{G}_1 \equiv \mc{G}$. 
A non-perturbative approximation can thus be obtained by a truncation to the $n$th order  of the $k$th hierarchy  of the time-local generator, e.g., 
\be
 \mc{G}_k \approx \sum_{m=1}^n \mc{G}_k^{(m)}
\ee  
\Eq{eq:104} then form a closed set of equations. The formal solution of $\mc{G}$ can be written as   
\be
\mc{G}(t) = \mc{G}^{(1)} \tord \exp\del{ \mc{G}_2^{(1)}  \tord \exp\del{ \mc{G}_3^{(1)} \tord \exp\del{ \cdots } }}
\ee 
In essence, the time-local generator is approximated by a nested time-ordered exponential function.

\section{Summary} \label{sec:summary}

To summarize, we have provided a formal proof of the diagrammatic representation of the generator of the  time-convolutionless master equation for an arbitrary open quantum system.  A nonperturbative approximation is developed based on calculating the time-local generator of the time-local generator repeatedly.  The time-local generator for the dynamical map is  expressed by a nested time-ordered exponential. 

In modeling the nonlinear  spectroscopy of an open quantum system, the multi-point time-correlation function are required. The two-point correaltion function can be computed as  
\be
C_{AB}(t) = \tr\cbr{ A(t) B \rho_0} = \tr\cbr{ A  \mc{U}\del{t, t_0}   {B\rho_0 }} = \tr_\text{S}\cbr{A \Lambda_{t,t_0}  B\rho_\text{S}(t_0)}
\ee 
where $\mc{U}(t, t_0)$ is the total propagator, and $A, B$ are system operators. 
%
For $n >2$-point correlation functions, .e.g., 
\be 
C(t_2, t_1) = \tr\cbr{ A(t_2) B(t_1)C \rho_0}  = \tr\cbr{A \mc{U}(t_2, t_1) B \mc{U}(t_1, t_0) C\rho_0}
,\ee
 this involves applying the $C$ operator to the initial density matrix, propagate for time duration $t_1$ using $\Lambda_{t,t_0}$, and then applying the $B$ operator followed by propagation of time duration $t_2 - t_1$. The problem is that since the total density matrix will become correlated (i.e., not a product state) after the first propagation so that one cannot use $\Lambda$ for the following propagation, i.e., 
\be \tr\cbr{A \mc{U}(t_2, t_1) B \mc{U}(t_1, t_0) C\rho_0} \ne \tr_\text{S}\cbr{A \Lambda(t_2, t_1) B \Lambda_{t_1, t_0} C \rho_\text{S}(t_0)}
\ee unless the system and bath remains uncorrelated for all times. 
The difficulty of computing multi-point correlation functions implies that getting the time-local generator is not enough to obtain all correlation functions of the system.

\begin{acknowledgments}
	We thank Dr. Yao Wang for helpful discussions.
\end{acknowledgments}

	
	\bibliography{../cavity,../optics,../OQS,../control,../dynamics}

\begin{thebibliography}{24}%
\makeatletter
\providecommand \@ifxundefined [1]{%
 \@ifx{#1\undefined}
}%
\providecommand \@ifnum [1]{%
 \ifnum #1\expandafter \@firstoftwo
 \else \expandafter \@secondoftwo
 \fi
}%
\providecommand \@ifx [1]{%
 \ifx #1\expandafter \@firstoftwo
 \else \expandafter \@secondoftwo
 \fi
}%
\providecommand \natexlab [1]{#1}%
\providecommand \enquote  [1]{``#1''}%
\providecommand \bibnamefont  [1]{#1}%
\providecommand \bibfnamefont [1]{#1}%
\providecommand \citenamefont [1]{#1}%
\providecommand \href@noop [0]{\@secondoftwo}%
\providecommand \href [0]{\begingroup \@sanitize@url \@href}%
\providecommand \@href[1]{\@@startlink{#1}\@@href}%
\providecommand \@@href[1]{\endgroup#1\@@endlink}%
\providecommand \@sanitize@url [0]{\catcode `\\12\catcode `\$12\catcode
  `\&12\catcode `\#12\catcode `\^12\catcode `\_12\catcode `\%12\relax}%
\providecommand \@@startlink[1]{}%
\providecommand \@@endlink[0]{}%
\providecommand \url  [0]{\begingroup\@sanitize@url \@url }%
\providecommand \@url [1]{\endgroup\@href {#1}{\urlprefix }}%
\providecommand \urlprefix  [0]{URL }%
\providecommand \Eprint [0]{\href }%
\providecommand \doibase [0]{http://dx.doi.org/}%
\providecommand \selectlanguage [0]{\@gobble}%
\providecommand \bibinfo  [0]{\@secondoftwo}%
\providecommand \bibfield  [0]{\@secondoftwo}%
\providecommand \translation [1]{[#1]}%
\providecommand \BibitemOpen [0]{}%
\providecommand \bibitemStop [0]{}%
\providecommand \bibitemNoStop [0]{.\EOS\space}%
\providecommand \EOS [0]{\spacefactor3000\relax}%
\providecommand \BibitemShut  [1]{\csname bibitem#1\endcsname}%
\let\auto@bib@innerbib\@empty
\bibitem [{\citenamefont {Breuer}\ \emph {et~al.}(2002)\citenamefont {Breuer},
  \citenamefont {Ma},\ and\ \citenamefont {Petruccione}}]{breuer2002}%
  \BibitemOpen
  \bibfield  {author} {\bibinfo {author} {\bibfnamefont {H.-P.}\ \bibnamefont
  {Breuer}}, \bibinfo {author} {\bibfnamefont {A.}~\bibnamefont {Ma}}, \ and\
  \bibinfo {author} {\bibfnamefont {F.}~\bibnamefont {Petruccione}},\
  }\href@noop {} {\bibfield  {journal} {\bibinfo  {journal}
  {arXiv:quant-ph/0209153}\ } (\bibinfo {year} {2002})},\ \Eprint
  {http://arxiv.org/abs/quant-ph/0209153} {arxiv:quant-ph/0209153} \BibitemShut
  {NoStop}%
\bibitem [{\citenamefont {Joos}\ \emph {et~al.}(2013)\citenamefont {Joos},
  \citenamefont {Zeh}, \citenamefont {Kiefer}, \citenamefont {Giulini},
  \citenamefont {Kupsch},\ and\ \citenamefont {Stamatescu}}]{Joos2013}%
  \BibitemOpen
  \bibfield  {author} {\bibinfo {author} {\bibfnamefont {E.}~\bibnamefont
  {Joos}}, \bibinfo {author} {\bibfnamefont {H.~D.}\ \bibnamefont {Zeh}},
  \bibinfo {author} {\bibfnamefont {C.}~\bibnamefont {Kiefer}}, \bibinfo
  {author} {\bibfnamefont {D.~J.}\ \bibnamefont {Giulini}}, \bibinfo {author}
  {\bibfnamefont {J.}~\bibnamefont {Kupsch}}, \ and\ \bibinfo {author}
  {\bibfnamefont {I.-O.}\ \bibnamefont {Stamatescu}},\ }\href@noop {} {\emph
  {\bibinfo {title} {Decoherence and the Appearance of a Classical World in
  Quantum Theory}}}\ (\bibinfo  {publisher} {{Springer Science \& Business
  Media}},\ \bibinfo {year} {2013})\BibitemShut {NoStop}%
\bibitem [{\citenamefont {Schlosshauer}(2007)}]{schlosshauer2007}%
  \BibitemOpen
  \bibfield  {author} {\bibinfo {author} {\bibfnamefont {M.~A.}\ \bibnamefont
  {Schlosshauer}},\ }\href@noop {} {\emph {\bibinfo {title} {Decoherence:
  {{And}} the {{Quantum-To-Classical Transition}}}}}\ (\bibinfo  {publisher}
  {{Springer Science \& Business Media}},\ \bibinfo {year} {2007})\BibitemShut
  {NoStop}%
\bibitem [{\citenamefont {Gu}\ and\ \citenamefont {Franco}(2017)}]{gu2017}%
  \BibitemOpen
  \bibfield  {author} {\bibinfo {author} {\bibfnamefont {B.}~\bibnamefont
  {Gu}}\ and\ \bibinfo {author} {\bibfnamefont {I.}~\bibnamefont {Franco}},\
  }\href {\doibase 10.1021/acs.jpclett.7b01817} {\bibfield  {journal} {\bibinfo
   {journal} {J. Phys. Chem. Lett.}\ }\textbf {\bibinfo {volume} {8}},\
  \bibinfo {pages} {4289} (\bibinfo {year} {2017})}\BibitemShut {NoStop}%
\bibitem [{\citenamefont {Gu}\ and\ \citenamefont {Franco}(2018)}]{gu2018}%
  \BibitemOpen
  \bibfield  {author} {\bibinfo {author} {\bibfnamefont {B.}~\bibnamefont
  {Gu}}\ and\ \bibinfo {author} {\bibfnamefont {I.}~\bibnamefont {Franco}},\
  }\href {\doibase 10.1103/PhysRevA.98.063412} {\bibfield  {journal} {\bibinfo
  {journal} {Physical Review A}\ }\textbf {\bibinfo {volume} {98}},\ \bibinfo
  {pages} {063412} (\bibinfo {year} {2018})}\BibitemShut {NoStop}%
\bibitem [{\citenamefont {Nielsen}\ and\ \citenamefont
  {Chuang}(2011)}]{Nielsen2011}%
  \BibitemOpen
  \bibfield  {author} {\bibinfo {author} {\bibfnamefont {M.~A.}\ \bibnamefont
  {Nielsen}}\ and\ \bibinfo {author} {\bibfnamefont {I.~L.}\ \bibnamefont
  {Chuang}},\ }\href@noop {} {\emph {\bibinfo {title} {Quantum {{Computation}}
  and {{Quantum Information}}: 10th {{Anniversary Edition}}}}},\ \bibinfo
  {edition} {10th}\ ed.\ (\bibinfo  {publisher} {{Cambridge University
  Press}},\ \bibinfo {address} {{New York, NY, USA}},\ \bibinfo {year}
  {2011})\BibitemShut {NoStop}%
\bibitem [{\citenamefont {Mukamel}(1995)}]{mukamel1995}%
  \BibitemOpen
  \bibfield  {author} {\bibinfo {author} {\bibfnamefont {S.}~\bibnamefont
  {Mukamel}},\ }\href@noop {} {\emph {\bibinfo {title} {Principles of Nonlinear
  Optical Spectroscopy}}}\ (\bibinfo  {publisher} {{Oxford University Press}},\
  \bibinfo {year} {1995})\BibitemShut {NoStop}%
\bibitem [{\citenamefont {Gu}\ and\ \citenamefont {Mukamel}(2020)}]{gu2020c}%
  \BibitemOpen
  \bibfield  {author} {\bibinfo {author} {\bibfnamefont {B.}~\bibnamefont
  {Gu}}\ and\ \bibinfo {author} {\bibfnamefont {S.}~\bibnamefont {Mukamel}},\
  }\href {\doibase 10.1039/C9SC04992D} {\bibfield  {journal} {\bibinfo
  {journal} {Chem. Sci.}\ }\textbf {\bibinfo {volume} {11}},\ \bibinfo {pages}
  {1290} (\bibinfo {year} {2020})}\BibitemShut {NoStop}%
\bibitem [{\citenamefont {Jin}\ \emph {et~al.}(2008)\citenamefont {Jin},
  \citenamefont {Zheng},\ and\ \citenamefont {Yan}}]{jin2008}%
  \BibitemOpen
  \bibfield  {author} {\bibinfo {author} {\bibfnamefont {J.}~\bibnamefont
  {Jin}}, \bibinfo {author} {\bibfnamefont {X.}~\bibnamefont {Zheng}}, \ and\
  \bibinfo {author} {\bibfnamefont {Y.}~\bibnamefont {Yan}},\ }\href {\doibase
  10.1063/1.2938087} {\bibfield  {journal} {\bibinfo  {journal} {J. Chem.
  Phys.}\ }\textbf {\bibinfo {volume} {128}},\ \bibinfo {pages} {234703}
  (\bibinfo {year} {2008})}\BibitemShut {NoStop}%
\bibitem [{\citenamefont {Nitzan}(2006)}]{nitzan2006}%
  \BibitemOpen
  \bibfield  {author} {\bibinfo {author} {\bibfnamefont {A.}~\bibnamefont
  {Nitzan}},\ }\href@noop {} {\emph {\bibinfo {title} {Chemical {{Dynamics}} in
  {{Condensed Phases}}: {{Relaxation}}, {{Transfer}} and {{Reactions}} in
  {{Condensed Molecular Systems}}}}}\ (\bibinfo  {publisher} {{Oxford
  University Press}},\ \bibinfo {year} {2006})\BibitemShut {NoStop}%
\bibitem [{\citenamefont {Scholes}\ \emph {et~al.}(2017)\citenamefont
  {Scholes}, \citenamefont {Fleming}, \citenamefont {Chen}, \citenamefont
  {{Aspuru-Guzik}}, \citenamefont {Buchleitner}, \citenamefont {Coker},
  \citenamefont {Engel}, \citenamefont {van Grondelle}, \citenamefont
  {Ishizaki}, \citenamefont {Jonas}, \citenamefont {Lundeen}, \citenamefont
  {McCusker}, \citenamefont {Mukamel}, \citenamefont {Ogilvie}, \citenamefont
  {{Olaya-Castro}}, \citenamefont {Ratner}, \citenamefont {Spano},
  \citenamefont {Whaley},\ and\ \citenamefont {Zhu}}]{scholes2017}%
  \BibitemOpen
  \bibfield  {author} {\bibinfo {author} {\bibfnamefont {G.~D.}\ \bibnamefont
  {Scholes}}, \bibinfo {author} {\bibfnamefont {G.~R.}\ \bibnamefont
  {Fleming}}, \bibinfo {author} {\bibfnamefont {L.~X.}\ \bibnamefont {Chen}},
  \bibinfo {author} {\bibfnamefont {A.}~\bibnamefont {{Aspuru-Guzik}}},
  \bibinfo {author} {\bibfnamefont {A.}~\bibnamefont {Buchleitner}}, \bibinfo
  {author} {\bibfnamefont {D.~F.}\ \bibnamefont {Coker}}, \bibinfo {author}
  {\bibfnamefont {G.~S.}\ \bibnamefont {Engel}}, \bibinfo {author}
  {\bibfnamefont {R.}~\bibnamefont {van Grondelle}}, \bibinfo {author}
  {\bibfnamefont {A.}~\bibnamefont {Ishizaki}}, \bibinfo {author}
  {\bibfnamefont {D.~M.}\ \bibnamefont {Jonas}}, \bibinfo {author}
  {\bibfnamefont {J.~S.}\ \bibnamefont {Lundeen}}, \bibinfo {author}
  {\bibfnamefont {J.~K.}\ \bibnamefont {McCusker}}, \bibinfo {author}
  {\bibfnamefont {S.}~\bibnamefont {Mukamel}}, \bibinfo {author} {\bibfnamefont
  {J.~P.}\ \bibnamefont {Ogilvie}}, \bibinfo {author} {\bibfnamefont
  {A.}~\bibnamefont {{Olaya-Castro}}}, \bibinfo {author} {\bibfnamefont
  {M.~A.}\ \bibnamefont {Ratner}}, \bibinfo {author} {\bibfnamefont {F.~C.}\
  \bibnamefont {Spano}}, \bibinfo {author} {\bibfnamefont {K.~B.}\ \bibnamefont
  {Whaley}}, \ and\ \bibinfo {author} {\bibfnamefont {X.}~\bibnamefont {Zhu}},\
  }\href@noop {} {\bibfield  {journal} {\bibinfo  {journal} {Nature}\ }\textbf
  {\bibinfo {volume} {543}},\ \bibinfo {pages} {647} (\bibinfo {year}
  {2017})}\BibitemShut {NoStop}%
\bibitem [{\citenamefont {Shapiro}\ and\ \citenamefont
  {Brumer}(2003)}]{shapiro2003}%
  \BibitemOpen
  \bibfield  {author} {\bibinfo {author} {\bibfnamefont {M.}~\bibnamefont
  {Shapiro}}\ and\ \bibinfo {author} {\bibfnamefont {P.}~\bibnamefont
  {Brumer}},\ }\href@noop {} {\emph {\bibinfo {title} {Principles of the
  {{Quantum Control}} of {{Molecular Processes}}}}}\ (\bibinfo  {publisher}
  {{Wiley-Interscience}},\ \bibinfo {year} {2003})\BibitemShut {NoStop}%
\bibitem [{\citenamefont {Breuer}\ \emph {et~al.}(2016)\citenamefont {Breuer},
  \citenamefont {Laine}, \citenamefont {Piilo},\ and\ \citenamefont
  {Vacchini}}]{breuer2016}%
  \BibitemOpen
  \bibfield  {author} {\bibinfo {author} {\bibfnamefont {H.-P.}\ \bibnamefont
  {Breuer}}, \bibinfo {author} {\bibfnamefont {E.-M.}\ \bibnamefont {Laine}},
  \bibinfo {author} {\bibfnamefont {J.}~\bibnamefont {Piilo}}, \ and\ \bibinfo
  {author} {\bibfnamefont {B.}~\bibnamefont {Vacchini}},\ }\href {\doibase
  10.1103/RevModPhys.88.021002} {\bibfield  {journal} {\bibinfo  {journal}
  {Rev. Mod. Phys.}\ }\textbf {\bibinfo {volume} {88}},\ \bibinfo {pages}
  {021002} (\bibinfo {year} {2016})}\BibitemShut {NoStop}%
\bibitem [{\citenamefont {Smirne}\ and\ \citenamefont
  {Vacchini}(2010)}]{smirne2010}%
  \BibitemOpen
  \bibfield  {author} {\bibinfo {author} {\bibfnamefont {A.}~\bibnamefont
  {Smirne}}\ and\ \bibinfo {author} {\bibfnamefont {B.}~\bibnamefont
  {Vacchini}},\ }\href {\doibase 10.1103/PhysRevA.82.022110} {\bibfield
  {journal} {\bibinfo  {journal} {Phys. Rev. A}\ }\textbf {\bibinfo {volume}
  {82}},\ \bibinfo {pages} {022110} (\bibinfo {year} {2010})}\BibitemShut
  {NoStop}%
\bibitem [{\citenamefont {Xu}\ \emph {et~al.}(2018)\citenamefont {Xu},
  \citenamefont {Yan}, \citenamefont {Liu},\ and\ \citenamefont
  {Shi}}]{xu2018}%
  \BibitemOpen
  \bibfield  {author} {\bibinfo {author} {\bibfnamefont {M.}~\bibnamefont
  {Xu}}, \bibinfo {author} {\bibfnamefont {Y.}~\bibnamefont {Yan}}, \bibinfo
  {author} {\bibfnamefont {Y.}~\bibnamefont {Liu}}, \ and\ \bibinfo {author}
  {\bibfnamefont {Q.}~\bibnamefont {Shi}},\ }\href {\doibase 10.1063/1.5022761}
  {\bibfield  {journal} {\bibinfo  {journal} {The Journal of Chemical Physics}\
  }\textbf {\bibinfo {volume} {148}},\ \bibinfo {pages} {164101} (\bibinfo
  {year} {2018})}\BibitemShut {NoStop}%
\bibitem [{\citenamefont {Breuer}\ \emph {et~al.}(2001)\citenamefont {Breuer},
  \citenamefont {Kappler},\ and\ \citenamefont {Petruccione}}]{breuer2001}%
  \BibitemOpen
  \bibfield  {author} {\bibinfo {author} {\bibfnamefont {H.-P.}\ \bibnamefont
  {Breuer}}, \bibinfo {author} {\bibfnamefont {B.}~\bibnamefont {Kappler}}, \
  and\ \bibinfo {author} {\bibfnamefont {F.}~\bibnamefont {Petruccione}},\
  }\href {\doibase 10.1006/aphy.2001.6152} {\bibfield  {journal} {\bibinfo
  {journal} {Ann. Phys.}\ }\textbf {\bibinfo {volume} {291}},\ \bibinfo {pages}
  {36} (\bibinfo {year} {2001})}\BibitemShut {NoStop}%
\bibitem [{\citenamefont {Kidon}\ \emph {et~al.}(2015)\citenamefont {Kidon},
  \citenamefont {Wilner},\ and\ \citenamefont {Rabani}}]{kidon2015}%
  \BibitemOpen
  \bibfield  {author} {\bibinfo {author} {\bibfnamefont {L.}~\bibnamefont
  {Kidon}}, \bibinfo {author} {\bibfnamefont {E.~Y.}\ \bibnamefont {Wilner}}, \
  and\ \bibinfo {author} {\bibfnamefont {E.}~\bibnamefont {Rabani}},\ }\href
  {\doibase 10.1063/1.4937396} {\bibfield  {journal} {\bibinfo  {journal} {J.
  Chem. Phys.}\ }\textbf {\bibinfo {volume} {143}},\ \bibinfo {pages} {234110}
  (\bibinfo {year} {2015})}\BibitemShut {NoStop}%
\bibitem [{\citenamefont {Tanimura}(2020)}]{tanimura2020}%
  \BibitemOpen
  \bibfield  {author} {\bibinfo {author} {\bibfnamefont {Y.}~\bibnamefont
  {Tanimura}},\ }\href {\doibase 10.1063/5.0011599} {\bibfield  {journal}
  {\bibinfo  {journal} {J. Chem. Phys.}\ }\textbf {\bibinfo {volume} {153}},\
  \bibinfo {pages} {020901} (\bibinfo {year} {2020})},\ \Eprint
  {http://arxiv.org/abs/2006.05501} {arxiv:2006.05501 [cond-mat,
  physics:physics, physics:quant-ph]} \BibitemShut {NoStop}%
\bibitem [{\citenamefont {Makri}\ and\ \citenamefont
  {Makarov}(1998)}]{makri1998}%
  \BibitemOpen
  \bibfield  {author} {\bibinfo {author} {\bibfnamefont {N.}~\bibnamefont
  {Makri}}\ and\ \bibinfo {author} {\bibfnamefont {D.~E.}\ \bibnamefont
  {Makarov}},\ }\href {\doibase 10.1063/1.469509} {\bibfield  {journal}
  {\bibinfo  {journal} {The Journal of Chemical Physics}\ }\textbf {\bibinfo
  {volume} {102}},\ \bibinfo {pages} {4611} (\bibinfo {year}
  {1998})}\BibitemShut {NoStop}%
\bibitem [{\citenamefont {Yan}(2014)}]{yan2014}%
  \BibitemOpen
  \bibfield  {author} {\bibinfo {author} {\bibfnamefont {Y.}~\bibnamefont
  {Yan}},\ }\href {\doibase 10.1063/1.4863379} {\bibfield  {journal} {\bibinfo
  {journal} {J. Chem. Phys.}\ }\textbf {\bibinfo {volume} {140}},\ \bibinfo
  {pages} {054105} (\bibinfo {year} {2014})}\BibitemShut {NoStop}%
\bibitem [{\citenamefont {Timm}(2011)}]{timm2011}%
  \BibitemOpen
  \bibfield  {author} {\bibinfo {author} {\bibfnamefont {C.}~\bibnamefont
  {Timm}},\ }\href {\doibase 10.1103/PhysRevB.83.115416} {\bibfield  {journal}
  {\bibinfo  {journal} {Phys. Rev. B}\ }\textbf {\bibinfo {volume} {83}},\
  \bibinfo {pages} {115416} (\bibinfo {year} {2011})}\BibitemShut {NoStop}%
\bibitem [{\citenamefont {Gu}(2020)}]{gu2020e}%
  \BibitemOpen
  \bibfield  {author} {\bibinfo {author} {\bibfnamefont {B.}~\bibnamefont
  {Gu}},\ }\href {\doibase 10.1103/PhysRevA.101.012121} {\bibfield  {journal}
  {\bibinfo  {journal} {Phys. Rev. A}\ }\textbf {\bibinfo {volume} {101}},\
  \bibinfo {pages} {012121} (\bibinfo {year} {2020})}\BibitemShut {NoStop}%
\bibitem [{\citenamefont {Gasbarri}\ and\ \citenamefont
  {Ferialdi}(2018)}]{gasbarri2018}%
  \BibitemOpen
  \bibfield  {author} {\bibinfo {author} {\bibfnamefont {G.}~\bibnamefont
  {Gasbarri}}\ and\ \bibinfo {author} {\bibfnamefont {L.}~\bibnamefont
  {Ferialdi}},\ }\href {\doibase 10.1103/PhysRevA.97.022114} {\bibfield
  {journal} {\bibinfo  {journal} {Phys. Rev. A}\ }\textbf {\bibinfo {volume}
  {97}},\ \bibinfo {pages} {022114} (\bibinfo {year} {2018})}\BibitemShut
  {NoStop}%
\bibitem [{\citenamefont {Gu}\ \emph {et~al.}(2021)\citenamefont {Gu},
  \citenamefont {Nenov}, \citenamefont {Segatta}, \citenamefont {Garavelli},\
  and\ \citenamefont {Mukamel}}]{gu2021}%
  \BibitemOpen
  \bibfield  {author} {\bibinfo {author} {\bibfnamefont {B.}~\bibnamefont
  {Gu}}, \bibinfo {author} {\bibfnamefont {A.}~\bibnamefont {Nenov}}, \bibinfo
  {author} {\bibfnamefont {F.}~\bibnamefont {Segatta}}, \bibinfo {author}
  {\bibfnamefont {M.}~\bibnamefont {Garavelli}}, \ and\ \bibinfo {author}
  {\bibfnamefont {S.}~\bibnamefont {Mukamel}},\ }\href {\doibase
  10.1103/PhysRevLett.126.053201} {\bibfield  {journal} {\bibinfo  {journal}
  {Phys. Rev. Lett.}\ }\textbf {\bibinfo {volume} {126}},\ \bibinfo {pages}
  {053201} (\bibinfo {year} {2021})}\BibitemShut {NoStop}%
\end{thebibliography}%
	
\end{document}